\documentclass[aps,prd,preprint,superscriptaddress,nofootinbib]{revtex4-1}
\usepackage{comment}
\usepackage[utf8]{inputenc} 
\usepackage{graphicx,color,overpic,mathtools}
\usepackage{amsthm,amsmath,amssymb,hyperref,mathrsfs}
\usepackage{braket,bm,bbm,setspace}
\PassOptionsToPackage{normalem}{ulem}
\usepackage{ulem} 
\usepackage{physics}
\usepackage{float}
\usepackage[makeroom]{cancel}
\usepackage[english]{babel}
\usepackage{graphicx}
\usepackage{color}
\graphicspath{ {images/} }

\addto\captionsspanish{}
\hypersetup{
    colorlinks=true,
    pdfborder={0 0 0},
    linkcolor=blue,
    citecolor=blue,
    urlcolor=blue
}
\definecolor{rRGB}{RGB}{255, 0, 0}

\begin{document}
\vspace*{-2cm}
\vspace*{0.0cm}
\title{Inner-extremal regular black holes from pure gravity}

\author{Francesco Di Filippo}%
\email{francesco.difilippo@mff.cuni.cz}
\affiliation{%
 Institute of Theoretical Physics, Faculty of Mathematics and Physics,
Charles University, V Holešovičkách 2, 180 00 Prague 8, Czech Republic
}%

\author{Ivan Kol{\'a}{\v r}}%
\email{ivan.kolar@matfyz.cuni.cz}
\affiliation{%
 Institute of Theoretical Physics, Faculty of Mathematics and Physics,
Charles University, V Holešovičkách 2, 180 00 Prague 8, Czech Republic
}%

\author{David Kubizňák}%
\email{david.kubiznak@matfyz.cuni.cz}
\affiliation{%
 Institute of Theoretical Physics, Faculty of Mathematics and Physics,
Charles University, V Holešovičkách 2, 180 00 Prague 8, Czech Republic
}%

\begin{abstract}
Recently it was shown that essentially all regular black hole models constructed so far can be obtained as solutions of vacuum gravity equations, upon considering an infinite series of quasi-topological higher curvature corrections. Here we show that such a construction can be upgraded to yield regular black holes with vanishing inner horizon surface gravity. 
In four dimensions, such a condition is necessary for the absence of classical instabilities associated with mass inflation on the inner horizon. 
\end{abstract}

\maketitle

\section{Introduction}

Black holes hide, deep in their interior,  theoretical evidence of the limitation of general relativity. It is now well known that black holes in general relativity must contain a singularity that cannot be described by the theory itself \cite{Penrose:1964wq,Hawking:1970zqf,Senovilla:1998oua}.  This promotes black holes as one of the best testing beds of any theory of gravity (possibly rivaled by cosmology). The standard lore is that a full theory of {quantum gravity} should be able to describe the physics at the singularity. If we assume that such a description can be done (at least in an approximate way) using the standard tools of pseudo-Riemannian geometry, then we only have a handful of possibilities \cite{Carballo-Rubio:Geodesically,Pandora,Carballo-Rubio:Darkness}. Among such possibilities, regular black holes constitute a conservative resolution of the singularity problem in which the central singularity is replaced by a non-singular core within an inner horizon \cite{Bardeen1968,Borde:1994ai,Dymnikova:1992ux,Bonanno:2000ep,Hayward:2005gi,Frolov:2016pav,Franzin:2022wai,Platania:2023srt}. 

Recently, it was shown that regular black holes can arise as vacuum solutions in purely gravitational theories \cite{Bueno:2024dgm}.
{The construction employs an infinite series\footnote{See \cite{Konoplya:2024hfg} for recent comments on `convergence' of this series.} of quasi-topological gravities \cite{Oliva:2010eb, Myers:2010ru,
Dehghani:2011vu, Cisterna:2017umf, Bueno:2019ycr, Bueno:2022res, Moreno:2023rfl}. 
These are higher-curvature gravities with fine-tuned coupling constants that were constructed to provide very simple field equations} for spherically symmetric 
configurations. Interestingly,
such theories 
are sufficient to capture the EFT corrections to vacuum gravity; any gravitational theory whose Lagrangian is a function of metric and the Riemann tensor 
can be mapped order by order to a quasi-topological gravity by a field re-definition \cite{Bueno:2019ltp, Bueno:2019ycr, Bueno:2024dgm}.\footnote{{
Any higher curvature gravity that is not Lovelock will be plagued by ghosts on some backgrounds.
However, in our paper we strictly restrict to spherical symmetry for which the corresponding equations of  motion (by the definition of quasi-topological gravity) remain of second order and no ghosts are present. The same remains true for spherical perturbations of our  spacetimes. Note also that in the framework of EFT (which however is not enough to remove the curvature singularities as done in our paper), all undesirable dofs from any EFT of gravity can be removed, e.g. \cite{Figueras:2024bba}.}} 
While it is only possible to construct these theories in higher than 4 dimensions,  
the simplicity of the field equations makes quasi-topological gravities useful test-grounds for investigating
properties of gravity beyond general relativity that would be technically too difficult to study in a generic 4-dimensional theory.
It is therefore very interesting to note that several regular black holes geometries can be obtained without the need of exotic matter.\footnote{A widely used model for constructing regular black holes as solutions of a concrete theory is to consider magnetically charged black holes in a particular model of non-linear electrodynamics, e.g. \cite{Ayon-Beato:1998hmi, Fan:2016hvf}. 
}

On the other hand, one of the main issues with regular black holes is that their inner horizon is generically unstable. One source of instability is the so-called mass inflation that leads to an exponential blueshift of classical perturbation at the inner horizon \cite{PhysRevLett.63.1663,Carballo-Rubio:2018pmi,Carballo-Rubio:2021bpr,DiFilippo:2022qkl,Carballo-Rubio:2024dca}. It is possible to get rid of this instability if the regular black hole is ``inner extremal", i.e., if the inner horizon has a zero value of the surface gravity \cite{Carballo-Rubio:2022kad,Franzin:2022wai} (see however \cite{McMaken:2023uue} for the analysis of the semiclassical instability).
In this paper, we study the possibility of having an inner extremal regular black hole as a solution of a quasi-topological theory of gravity. 
{Namely, we show that it is possible to construct a theory that admits inner-extremal regular black holes as solutions and provide an explicit example of a regular black hole with triple degenerate inner horizon and one non-extremal outer horizon.}\footnote{{Very recently a `super-extremal' black hole with quadruple degenerate horizon in quasi-topological electromagnetism was constructed in \cite{Hod:2024shg}.}}
Unfortunately, {such a construction} 
is only possible if the mass of the black hole is tuned to have a specific relation with the parameters of the action. In other words, it is not possible to have inner extremal black holes with arbitrary mass in such a theory. We prove that this limitation is generic and shared by all quasi-topological theories of gravity.


\section{Tower of quasi-topological densities}
Let us start by summarizing the main properties of the theories containing quasi-topological densities.
Starting from a general higher-curvature theory 
of the form
${\cal L}(g^{ab}, R_{abcd})$, where $R_{abcd}$ is the Riemann tensor, the field equations take the following form:
\begin{equation}
{\cal E}_{ab}=P_a{}^{cde} R_{bcde}-\frac{1}{2}g_{ab}{\cal L} -2\nabla^c \nabla^d P_{acdb}=0\;,\quad P^{abcd}\equiv \frac{\partial {\cal L}}{\partial R_{abcd}}\;.
 \end{equation} 
Quasi-topological gravities are a special class of the above, for which \cite{Bueno:2019ycr, Bueno:2022res}
\begin{equation}
\nabla^d P_{acdb}=0    
\end{equation}
for general static spherically symmetric metrics. Such theories are non-trivial in $D\geq 5$ dimensions.

The generic action in ${D\geq 5}$ dimensions can be written as 
\begin{equation}\label{eq:action}
    I_{\mathrm{QT}}=\frac{1}{16 \pi G} \int \mathrm{d}^D x \sqrt{|g|}\left[R+\sum_{n=2}^{\infty} \alpha_n \mathcal{Z}_n\right]\;,
\end{equation}
where $\alpha_n$ are free parameters and $\mathcal{Z}_n$ are specific combinations of curvature scalars of order $n$, whose specific expressions can be found in \cite{Bueno:2024dgm}.
While at every order there exist several quasi-topological densities, one can choose one representative density ${\cal Z}_n$; the contribution of other densities to  the equations of motion for static spherically symmetric solutions is the same. 

For a static and spherically symmetric ansatz
\begin{equation}
    \mathrm{d} s^2=-N(r)^2 f(r) \mathrm{d} t^2+\frac{\mathrm{d} r^2}{f(r)}+r^2 \mathrm{~d} \Omega_{D-2}^2\;,
\end{equation}
the corresponding field equations take the following simple form:
\begin{equation}\label{eq:EOM}
    \frac{d N}{d r}=0\;, \quad \frac{d}{d r}\left[r^{D-1} h(\psi)\right]=0\;,
\end{equation}
where $h$ is an analytic function (at ${\psi=0}$) with the expansion given by parameters $\alpha_n$,
\begin{equation}
    h(\psi) \equiv \psi+\sum_{n=2}^{\infty} \alpha_n \psi^n, \quad \psi \equiv \frac{1-f(r)}{r^2}\;.
\end{equation}
The solution of Eqs. \eqref{eq:EOM} is {$N=1$ (upon normalizing the timelike Killing vector $\partial_t$),} and
\begin{equation}\label{eq:feq-algebraic}
    h(\psi)=\frac{m}{r^{D-1}}\;,
\end{equation}
where $m$ is  an integration constant which can be related to the mass of the spacetime. Since $h(\psi)$ is an analytic function of $\psi$, then its inverse, $\psi=h^{-1}(m/r^{D-1})$, should be an analytic function of $m/r^{D-1}$, which we denote by $\xi:=h^{-1}$. Consequently, the function $f$ should take the form 
\begin{equation}\label{eq:fanalytic}
    f=1-r^2\xi\left(\frac{m}{r^{D-1}}\right)\;,
\end{equation}
where $\xi$ is analytic at zero.

\section{Inner extremal solutions}
We now want to investigate if among the theories \eqref{eq:action} there are some that admit inner extremal regular black holes as their vacuum solutions. 

In order to satisfy the regularity conditions $f(0)=1$ and $f'(0)=0$, and the correct asymptotic behavior $f = 1-m/r^{{D-3}}+O(1/r^{D-2})$ for $r\to\infty$ we need to have two horizons where the inner one is triple degenerate (and the outer one is non-degenerate)~\cite{Carballo-Rubio:2022kad}, 
\begin{equation}
    f(r_{-})=f'(r_{-})=f''(r_{-})=0=f(r_{+}) \qquad \text{where} \qquad 0<r_{-}<r_{+}<\infty\;.
\end{equation}
Taking into account \eqref{eq:fanalytic}, we can start with the ansatz
\begin{equation}\label{eq:fr}
    f(r)=1-r^2\frac{A\left(\frac{m}{r^{D-1}}\right)}{B\left(\frac{m}{r^{D-1}}\right)}\;,
\end{equation}
where $A$ and $B$ are two quadratic functions
\begin{equation}
    A(x)=a_0+a_1 x+ a_2 x^2\;, \quad B(x)=b_0+b_1 x+ b_2 x^2\;.
\end{equation}
Assuming $a_2\neq0$ and $b_2\neq0$, the regularity condition $f(r)=1+\mathcal{O}(r^2)$ is automatically satisfied. In order to achieve the asymptotic behavior $f(r)=1-\frac{m}{r^{D-3}}+\mathcal{O}(\frac{1}{r^{D-2}})$ for $r\to\infty$, we will set $a_0=0$ and $a_1=b_0\neq 0$. Without loss of generality, we can set $b_0=1$, because dividing the numerator and denominator by $b_0$ will not affect the function. Hence the metric is parameterized by $a_2$, $b_1$, $b_2$, and $m$. Since only $m$ is the free parameter of the solution. The remaining three constants $a_2$, $b_1$, $b_2$ are parameters of the theory, which is chosen requiring that 
\begin{equation}\label{eq:hpsi}
\begin{aligned}
    h(\psi) &=\frac{1-b_1\psi-\sqrt{(1-b_1 \psi )^2-4  \psi  (b_2 \psi -a_2)}}{2 (b_2 \psi -a_2)} 
    \\
    &=\psi +(b_1-a_2)\psi ^2 +\left(-3 a_2 b_1+2 a_2^2+b_1^2+b_2\right)\psi ^3 
    \\
    &\quad+ \left(10 a_2^2 b_1-6 a_2 b_1^2-4 a_2 b_2-5 a_2^3+b_1^3+3 b_1 b_2\right) \psi ^4+\mathcal{O}\left(\psi ^5\right)\;
\end{aligned}
\end{equation}
is a solution.

The function $h(\psi)$ was obtained by inverting $\psi=\frac{w (a_2 w+1)}{b_2 w^2+b_1 w+1}$ as a function of the right-hand side of the algebraic field equation \eqref{eq:feq-algebraic}, $w:=m/r^{D-1}$, and dropping the root for which $h(0)\neq 0$. Recall that the coupling constants are given by $\alpha_n=h^{(n)}(0)/n!$; 
since $h(\psi)$ has an explicit closed form expression, it specifies the complete theory. 

Let us first show that this theory admits a 1-parametric solution with four non-degenerate positive roots $r_{i}$, $i=1,2,3,4$, i.e., $f(r_{i})=0$ with $f'(r_{i})\neq0$. This requires us to satisfy the following conditions
\begin{equation}
    1-\frac{m r_{i}^3 \left(a_2 m r_{i}+b_0 r_{i}^D\right)}{b_1 m r_{i}^{D+1}+b_0 r_{i}^{2 D}+b_2 m^2 r_{i}^2}=0\;, \quad  i=1,2,3,4\;.
\end{equation}
These four equations can be solved analytically with respect to $a_2$, $b_1$, $b_2$, and $m$ in any $D\geq5$. Since the final expression is a bit cumbersome, we will work in the simplest case, $D=5$. Then the result reads
\begin{equation}\label{eq:coeffd5}
\begin{aligned}
    a_2 &= \frac{
 r_1^2 r_2^2 r_3^2 + r_1^2r_2^2 r_4^2 +  
    r_1^2 r_3^2 r_4^2 +r_2^2 r_3^2 r_4^2 }{(r_1^2 + 
   r_2^2 + r_3^2 + r_4^2)^2}\;, 
   \\
b_1 &= \frac{
     r_1^2r_2^2 + r_1^2r_3^2 + r_1^2r_4^2 + r_2^2r_3^2 + r_2^2r_4^2 + r_3^2 r_4^2}{
 r_1^2 + r_2^2 + r_3^2 + r_4^2}\;, 
 \\
b_2 &= \frac{
  r_1^2 r_2^2 r_3^2 r_4^2}{(r_1^2 + r_2^2 + r_3^2 + 
   r_4^2)^2}\;, 
   \\
   m &= r_1^2 + r_2^2 + r_3^2 + r_4^2\;.
\end{aligned}
\end{equation}
After substituting this back to the ansatz, we arrive at
\begin{equation}
    f(r)=\frac{\left(r^2-r_1^2\right) \left(r^2-r_2^2\right) \left(r^2-r_3^2\right) \left(r^2-r_4^2\right)}{r^8+\left(r_1^2r_2^2 + r_1^2r_3^2 + r_1^2r_4^2 + r_2^2r_3^2 + r_2^2r_4^2 + r_3^2 r_4^2\right)r^4+r_1^2 r_2^2 r_3^2 r_4^2}\;.
\end{equation}
Consequently, given a theory described by three constants $a_2$, $b_1$, $b_2$, the above function $f(r)$ is a one-parametric class of solutions with the free parameter $m$. The constants $r_i$ are given implicitly as solutions of \eqref{eq:coeffd5}.

\begin{figure}
    \centering
    \includegraphics[width=0.8\columnwidth]{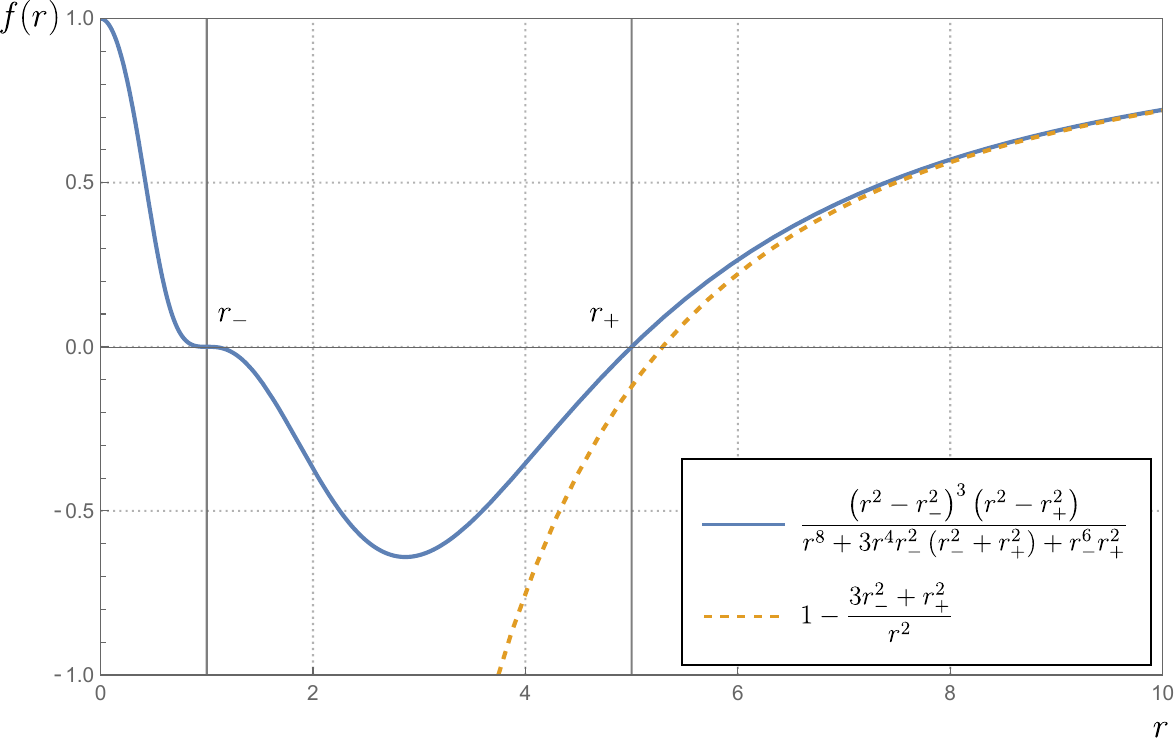}
    \caption{{\bf Metric function of inner extremal black hole.} We display $f(r)$ for the inner-extremal (blue solid curve) regular black hole in $D=5$ with $r_{-}=1$ and $r_{+}=5$. This is compared to $f(r)$ of 5D Schwarzschild--Tangherlini solution (orange dashed curve) for the same choice of mass parameter $m=3r_{-}^2+r_{+}^2$.}
    \label{fig:graph}
\end{figure}

Let us now consider the inner-extremal regular black hole. The expression for $f(r)$ can be obtained by taking the limit in which the three inner-most horizons coincide, $r_{-}\equiv r_1=r_2=r_3$, while the outer remains non-degenerate $r_{+}\equiv r_4$ so that $0<r_{-}<r_{+}$. (Alternatively, we could also solve $f(r_{-})=f'(r_{-})=f''(r_{-})=0$ and $f(r_{+})=0$ and obtain the same metric.) The solution takes the following form:
\begin{equation}\label{eq:frinnerextremed5}
    f(r) =\frac{\left(r^2-r_-^2\right)^3 \left(r^2-r_+^2\right)}{r^8+3 r^4 r_-^2 \left(r_-^2+r_+^2\right)+r_-^6 r_+^2}\;,
\end{equation}
where the two parameters $r_{+}$ and $r_{-}$ are given implicitly as the solutions of
\begin{equation}
    a_2=\frac{r_-^4 \left(r_-^2+3 r_+^2\right)}{\left(3 r_-^2+r_+^2\right){}^2}\;, \quad b_1=\frac{3 \left(r_-^4+r_+^2 r_-^2\right)}{3 r_-^2+r_+^2}\;, \quad b_2=\frac{r_-^6 r_+^2}{\left(3 r_-^2+r_+^2\right){}^2}\;, \quad m=3 r_-^2+r_+^2\;.
\end{equation}
This set of equations is over-determined, which reflects the fact that the original parameters $a_2$, $b_1$, $b_2$, $m$ are constrained by the conditions on coinciding roots. This has two important consequences: i) The solution only exists for the theory satisfying
\begin{equation}
    a_2=\frac{b_1^3+108 b_2 b_1-\left(b_1^2-36 b_2\right)^{\frac32} }{18 \left(b_1^2+12 b_2\right)}\;, \quad b_1 > 0 \;, \quad 0<b_2<\frac{b_1^2}{36}\;.
\end{equation}
Hence, there are only two free parameters of the theory, e.g., the parameters $b_1$ and $b_2$ (while $a_2$ is determined in terms of these). Furthermore, they are restricted by the inequalities coming from $0<r_{-}<r_{+}$. ii) More importantly, the parameter $m$ is no longer arbitrary, because the choice of $b_1$ and $b_2$ also fixes its value,
\begin{equation}
    m=\frac{b_1^3+108 b_2 b_1+\left(b_1^2-36 b_2\right)^{\frac32} }{54 b_2}\;.
\end{equation}
The positions of the horizons are given by
\begin{equation}
    r_{-}^2=\frac{1}{3} \left(2 b_1-\sqrt{b_1^2-36 b_2}\right)\;, \quad
    r_{+}^2=\frac{b_1^3+ \left(b_1^2+18 b_2\right)\sqrt{b_1^2-36 b_2}}{54 b_2}\;.
\end{equation}
The graph of the function \eqref{eq:frinnerextremed5} is shown in Fig.~\ref{fig:graph}.

To conclude this section, let us write the inner-extremal regular black hole in a generic number of dimensions $D$,
\begin{equation}
\begin{aligned}
    a_2 &=\tfrac{r_+ r_-^{D+3} \left(-\left((D+1) r_+^3 r_-^D\right)+(D-3) r_+ r_-^{D+2}+(D+1) r_-^3 r_+^D-(D-3) r_- r_+^{D+2}\right) }{\left(r_+^2 \left((D-1)^2 r_+^2-(D-3) (D-2) r_-^2\right) r_-^{2 D}-4 (D-2) r_+^{D+1} r_-^{D+3}+(D-3) r_-^4 r_+^{2 D}\right){}^2}
    \\
    &\quad\times \left(r_+^{D+1} \left((D-1)^2 r_+^2-(D-3) (D+1) r_-^2\right) r_-^D+(D-3) r_+^2 r_-^{2 D+1}-(D+1) r_-^3 r_+^{2 D}\right)\;,
    \\
    b_1 &=\tfrac{(D+1) r_-^6 r_+^{2 D}-4 (D-2) r_-^{D+3} r_+^{D+3}-(D-3) r_+^2 r_-^{2 D+2} \left((D-2) r_-^2-(D+1) r_+^2\right)}{r_+^2 \left((D-1)^2 r_+^2-(D-3) (D-2) r_-^2\right) r_-^{2 D}-4 (D-2) r_+^{D+1} r_-^{D+3}+(D-3) r_-^4 r_+^{2 D}}\;,
    \\
    b_2 &=\tfrac{(D-3) r_+ r_-^{D+4} \left(-\left((D+1) r_+^3 r_-^D\right)+(D-3) r_+ r_-^{D+2}+(D+1) r_-^3 r_+^D-(D-3) r_- r_+^{D+2}\right) }{\left(r_+^2 \left((D-1)^2 r_+^2-(D-3) (D-2) r_-^2\right) r_-^{2 D}-4 (D-2) r_+^{D+1} r_-^{D+3}+(D-3) r_-^4 r_+^{2 D}\right){}^2}
    \\
    &\quad\times\left(r_+^4 r_-^{2 D}-(D-2) \left(r_-^2-r_+^2\right) r_+^{D+1} r_-^{D+1}-r_-^4 r_+^{2 D}\right)\;,
    \\
    m &=\tfrac{-(D-1)^2 r_+^4 r_-^{2 D}+4 (D-2) r_+^{D+1} r_-^{D+3}+(D-3) (D-2) r_+^2 r_-^{2 D+2}-(D-3) r_-^4 r_+^{2 D}}{r_-^3 r_+ \left(-\left((D+1) r_+^3 r_-^D\right)+(D-3) r_+ r_-^{D+2}+(D+1) r_-^3 r_+^D-(D-3) r_- r_+^{D+2}\right)}\;.
\end{aligned}
\end{equation}
This is a solution of the theory given by \eqref{eq:hpsi}, where the parameters and mass are restricted by formulas analogous to the $D=5$ case discussed above. Unfortunately, these formulas cannot be written explicitly because expressing all constants only in terms of $b_1$ and $b_2$ would require solving higher-order polynomial equations.

\section{No degenerated horizons for arbitrary mass}
As we just observed, the degeneracy of the horizons implied that $m$ was no longer an arbitrary parameter, but it was determined by means of the parameters of the theory. Let us now show that this is a generic feature of all theories with the action \eqref{eq:action}. As mentioned above, the function $f$ should take the form $f=1-r^2\xi(m/r^4)$ where $\xi$ is an analytic function at zero; otherwise $h(\psi)$ could not be analytic at $\psi=0$. In order to have a degenerated horizon at $r=r_0$, we demand that
\begin{equation}
    \begin{aligned}
        0&=f(r_0)=1-r_0^2\xi(m/r_0^4)\;,
        \\
        0&=f'(r_0)=-2r_0\left[\xi(m/r_0^4)-2\xi'(m/r_0^4)(m/r_0^4)\right]\;.
    \end{aligned}
\end{equation}
We will show that these two conditions are inconsistent. 

Let us first take a look at the second equation ($r=r_0$ is a stationary point) and let us denote $\mu:=m/r_0^4$. If $\mu$ depends on $m$ (in general $r_0=r_0(m)$), then the second equation is a differential equation for $\xi(\mu)$ leading to $\xi\propto\sqrt{\mu}$. This corresponds to $f$ being independent of $r$ in some interval corresponding to the range of the variable $\mu$. Since a constant profile $f(r)$ in some interval of $r$ seems rather unphysical,\footnote{Focusing on analytic $f(r)$ the only option that is regular at the origin would be the flat spacetime.} one should require $\mu$ to be independent of $m$. Then the second equation is an equation relating values of $\xi$ and its derivative at a single point $\mu$ (i.e., it is not a differential equation) and it implies that $r_0$ should depend on mass as $r_0=(m/\mu)^{1/4}$. (Note that this is actually satisfied at the stationary points of all solutions presented in \cite{Bueno:2024dgm}.) 

If we now demand the first equation ($r=r_0$ is a horizon) to hold as well, then we find that $r_0$ has to be independent of $m$ because this equation implies $r_{0}=1/\sqrt{\xi(\mu)}$. This is clearly inconsistent with the previous dependence of $r_0$ on $m$ and means that we cannot maintain $m$ to be the free parameter in the solutions with degenerated horizons.

A simpler example of this behavior is obtained from the Hayward black hole, $f(r)=1- r^2/(r^4/m+\alpha)$, which is a solution of the theory given by ${\alpha_n=\alpha^{n-1}}$ for arbitrary value of $m$. The degeneration of the horizons, however, requires relating the mass $m$ to the parameter of the theory $\alpha$ by fixing its value to ${m=4\alpha}$. One can see that the degenerated root satisfies $r_0=\sqrt{2\alpha}=\sqrt{m/2}$, so $\mu=4/m$ is not independent of $m$.

\section{Summary}
In this short paper we have studied the solutions of a class of higher-dimensional quasi-topological theories of gravity and we have shown that it is possible to identify some theories within this class that admit inner extremal regular black holes as vacuum solutions. We have presented a simple example of such a theory -- it is given by an explicit infinite sum of quasitopological gravities. Whether such a series truly converges in a rigorous mathematical sense to a theory that ``preserves quasi-topological features"  remains to be seen.

The inner extremal condition is interesting as (in 4-dimensions) it provides a simple resolution of the problems associated to the classical instability of the inner horizon that spoils the viability of generic classes of regular black holes.  However, we have shown that inner extremal solutions only exist for a specific value of the mass. Thus, most compact objects described in this theory have multiple horizons rather than a degenerate inner one. We have proved that this issue is completely generic and it applies to every quasi-topological theory. This issue might be resolved upon considering  black holes with {simple matter fields.} For example, upon including Maxwell charge, the equation of motion \eqref{eq:feq-algebraic} gets simply replaced by 
\begin{equation}
    h(\psi)=\frac{m}{r^{D-1}}-\frac{q^2}{r^{2(D-2)}}\;.
\end{equation}
The corresponding inner-extremal black hole will likely preserve one free parameter. 

In conclusion, quasi-topological theories provide a simplified setting to study fundamental questions that would not be possible to address otherwise with the expectation that some of the lessons drawn will apply to more generic frameworks. For instance, future research could focus on the study of gravitational collapse in this theories that can shed some light on the dynamical formation of regular black holes.

\section*{Acknowlegments}

{We would like to thank Robie Hennigar and Tayebeh Tahamtan for useful comments on our work.}
F.D.F acknowledge financial support by Primus grant PRIMUS/23/SCI/005 from Charles University and by GAČR 23-07457S grant of the Czech Science Foundation.
I.K. acknowledge financial support by Primus grant PRIMUS/23/SCI/005 from Charles University.
D.K. acknowledge financial support from GAČR 23-07457S grant of the
Czech Science Foundation.
All three authors are grateful to the Charles University Research Center Grant
No. UNCE24/SCI/016 for their support.


%

\end{document}